\documentclass[aps,prb,preprint,superscriptaddress]{revtex4-2}

\usepackage{graphicx}
\usepackage{amsmath}
\usepackage{amssymb}
\usepackage{mathrsfs}
\usepackage[T1]{fontenc}
\usepackage{calligra}
\usepackage{array}
\usepackage{float}
\usepackage{bm}
\usepackage[toc,page]{appendix}
\usepackage{braket}
\usepackage{wrapfig}

\begin{document}


\title{Fluctuation Spectroscopy in Granular Superconductors with Application to Boron-doped Nanocrystalline Diamond}


\author{D. T. S. Perkins}
\email[]{DTP333@student.bham.ac.uk}
\affiliation{School of Physics and Astronomy, University of Birmingham, Edgbaston, Birmingham, B15 2TT, United Kingdom}

\author{G. M. Klemencic}
\affiliation{School of Physics and Astronomy, Cardiff University, Queen's Building, The Parade, Cardiff, CF24 3AA, United Kingdom}

\author{J. M. Fellows}
\affiliation{School of Physics, H. H. Wills Physics Laboratory, University of Bristol, Tyndall Avenue, Bristol, BS8 1TL, United Kingdom}

\author{R. A. Smith}
\affiliation{School of Physics and Astronomy, University of Birmingham, Edgbaston, Birmingham, B15 2TT, United Kingdom}


\date{\today}

\begin{abstract}
	We perform a detailed calculation of the various contributions to the fluctuation conductivity of a granular metal close to its superconducting transition.
	We find three distinct regions of power law behavior in reduced temperature, $\eta=(T-T_c)/T_c$, with crossovers at $\Gamma/T_c$ and $E_{Th}/T_c$, where $\Gamma$ is the electron tunneling rate, and $E_{Th}$ is the Thouless energy of a grain. The calculation includes both intergrain and intragrain degrees of freedom. This complete theory of the fluctuation region in granular superconductors is then compared to experimental results from boron-doped nanocrystalline diamond, using the assumption of a constant phase breaking rate, $\tau_{\phi}^{-1}$. We find a semi-quantitative agreement between the theoretical and experimental results only in the case of large phase breaking. We argue that there may be a novel phase breaking mechanism in granular metals worthy of further experimental and theoretical investigation.
\end{abstract}


\maketitle

\section{Introduction} \label{Sec_intro}
Electronic transport in granular materials has been studied both theoretically and experimentally in great depth since the 1960s, with earlier works focussing heavily on the behavior near the metal insulator transition \cite{Abeles1975,Edwards_Rao_book}. In the late 1990s and early 2000s, theoretical developments allowed for calculation of transport properties of granular materials in the metallic regime; see \cite{Beloborodov2007} for a review. In these works a granular diagrammatic theory was used to determine the effects of weak localization \cite{Beloborodov2004,Biagini_2005_WL}, electron-electron interactions \cite{Beloborodov2001,Beloborodov2003,Efetov2003}, and superconducting fluctuations in the presence of magnetic fields \cite{Beloborodov1999,Beloborodov2000,Skrzynski_2002} on the electrical conductivity of a granular metal. This approach only considered the intergrain degrees of freedom (DOFs), and generally led to the same temperature dependences seen in homogeneous materials, albeit with a change to the relevant physical parameters, such as the effective diffusion constant.

In contrast, Lerner et. al. \cite{Lerner} considered superconducting fluctuations in granular metals from the perspective of the intragrain DOFs only. They predicted two crossovers in the temperature dependence of the fluctuation conductivity, $\sigma_{fl}$, as one moves away from the superconducting transition temperature, $T_{c}$. They assumed that, for a typical granular metal, $\delta \ll \Gamma \lesssim E_{Th} \lesssim T_{c}$, where $\delta$ is the mean level spacing, $\Gamma$ is the tunneling rate of electrons between grains, $E_{Th} = \mathcal{D}_{0}/a^{2}$ is the Thouless energy of a single grain, $\mathcal{D}_{0}$ is the intragrain diffusion coefficient, and $a$ is the typical grain size. This assumption  means that the system is in the metallic limit with the dimensionless tunneling conductance $g_{T} = \Gamma/\delta \gg 1$. The granular nature of the system is ensured since $g_{g} = E_{Th}/\delta \gtrsim g_{T}$, where $g_{g}$ is the dimensionless conductance of an isolated grain. Lerner et. al. \cite{Lerner} predicted the first of these crossovers to occur as the reduced temperature, $\eta = (T-T_{c})/T_{c}$, approached $\Gamma/T_{c}$, and the second to happen at $E_{Th}/T_{c}$.

Two recent papers by Klemencic et. al. \cite{Klemencic2017,Klemencic2019} characterised the granular transport properties in films of boron-doped nanocrystalline diamond (BNCD), where $T_{c} \lesssim 4$K. In \cite{Klemencic2017} they measured the corrections to the electrical conductivity due to superconducting fluctuations and observed two crossovers in the behavior of the fluctuation conductivity. As $T$ increased, the power law behavior, $\sigma_{fl} \sim \eta^{\alpha}$, changed from $\alpha = -1/2$ (\textit{close-to-}$T_{c}$ \textit{region}), to $\alpha = -3$ (\textit{intermediate region}), and back to $\alpha = -1/2$ (\textit{far-from-}$T_{c}$ \textit{region}). This matched Lerner et. al.'s prediction for the close-to-$T_{c}$ and intermediate regions, but not the far-from-$T_{c}$ region, where they predicted a power law with $\alpha = -2$.

In this paper we combine the approaches of Beloborodov et. al. \cite{Beloborodov2007} and Lerner et. al. \cite{Lerner} to consider both internal (intragrain) and external (intergrain) DOFs simultaneously. This enables us to consistently treat the various temperature regimes: the external DOFs are needed to consider the close-to-$T_{c}$ to intermediate crossover; the internal DOFs are needed to consider the intermediate to far-from-$T_{c}$ crossover. We first tackle the problem analytically for a general granular metal, and obtain power laws for the different contributions to $\sigma_{fl}$ in each region. We then use the material parameters obtained by Klemencic et. al. to numerically evaluate the predictions in BNCD for comparison to experiment.

The rest of the paper is structured as follows: Section \ref{Granular_diagrammatics_sec} outlines and extends the granular diagrammatic formalism to include both internal and external DOFs. The three regions of behavior in $\sigma_{fl}$ emerge naturally, with simple power law relations appearing deep inside each region due to the changing energy scale set by the pole of the pair propagator.

In Section \ref{Corrections_sec} we perform the main diagrammatic calculation, which includes both internal and external DOFs. We analytically consider the limiting power law behavior of the Aslamazov-Larkin (AL), Maki-Thompson (MT), and density of states (DOS) contributions in each region, and summarise these in table \ref{results_table}. In Section \ref{Theoretical_discussion_sec} we discuss the different power laws produced by each contribution in each region, and their relative sizes. This allows us to make experimental predictions for superconducting fluctuations in granular metals. In Section \ref{Comparison_sec} we compare our theory to the experimental measurements of fluctuation conductivity in BNCD by Klemencic et. al. \cite{Klemencic2017}. We find that the inclusion of phase breaking is necessary to obtain semi-quantitative agreement with experiment.

\section{Diagrammatic Theory for Granular Systems} \label{Granular_diagrammatics_sec}

We assume a very general form for the Hamiltonian of a granular system with superconducting correlations,
\begin{equation}
\begin{split}
	H &= \sum_{i}\sum_{\sigma}\sum_{\mathbf{k}} \xi_{\mathbf{k}}^{\null} c_{i\sigma\mathbf{k}}^{\dagger}c_{i\sigma\mathbf{k}}^{\null} + \sum_{i}\sum_{\sigma}\sum_{\mathbf{k},\mathbf{q}} U_{i}(\mathbf{q}) c_{i\sigma\mathbf{k}+\mathbf{q}}^{\dagger}c_{i\sigma\mathbf{k}}^{\null} + \sum_{i,j}\sum_{\sigma}\sum_{\mathbf{k},\mathbf{p}} t_{ij}^{\mathbf{k}\mathbf{p}} c_{i\sigma\mathbf{k}}^{\dagger}c_{j\sigma\mathbf{p}}^{\null} \\
	&\quad+ \frac{1}{2}\sum_{i,j}\sum_{\sigma,\sigma'}\sum_{\mathbf{k},\mathbf{p}} V_{ij} c_{i\sigma\mathbf{k}}^{\dagger}c_{j\sigma'\mathbf{p}}^{\dagger} c_{j\sigma'\mathbf{p}}^{\null}c_{i\sigma\mathbf{k}}^{\null} - \lambda\sum_{i}\sum_{\substack{\mathbf{k},\mathbf{p}, \\ \mathbf{q}}} c_{i\uparrow\mathbf{k}}^{\dagger}c_{i\downarrow\mathbf{q}-\mathbf{k}}^{\dagger} c_{i\downarrow\mathbf{q}-\mathbf{p}}^{\null}c_{i\uparrow\mathbf{p}}^{\null}.
	\label{Hamiltonian}
\end{split}
\end{equation}
The first term is the single grain free-electron Hamiltonian, where $\xi_{\mathbf{k}} = k^{2}/(2m) - \mu$ is the electron energy relative to the Fermi surface; the second term describes the random scattering from impurities in a single grain; the third term accounts for tunneling between grains with the matrix element $t_{ij}^{\mathbf{k}\mathbf{p}}$ being associated to tunneling from state $\mathbf{p}$ in the $j$\textsuperscript{th} grain to state $\mathbf{k}$ in the $i$\textsuperscript{th} grain. The last two terms describe electron-electron interactions. The fourth term is the Coulomb repulsion, $V_{ij}$, where we neglect dependence upon intragranular momenta, $q \sim a^{-1}$, since these are much larger than the intergranular momenta, $Q \sim L^{-1}$, where $L = \mathcal{N}^{1/3}a$ is the typical system length, and $\mathcal{N}$ is the number of grains. Finally, the fifth term is the standard s-wave BCS interaction. We may assume that the BCS interaction acts within a grain, so that our system forms a Josephson junction array for $T < T_{c}$.

\begin{figure}[t]
	\centering
	\includegraphics[width=8.6cm]{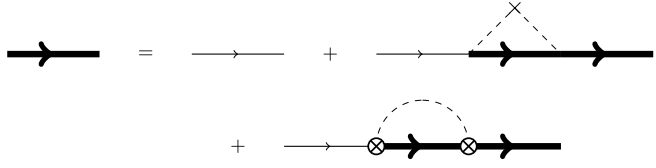}
	\caption{Dyson equation for the granular electron Green's function.}
	\label{electron_GF_diagram}
\end{figure}

In this paper we focus on the corrections to the electrical conductivity due to superconducting fluctuations. Our methodology differs from the previous granular diagrammatic literature \cite{Beloborodov1999,Beloborodov2000,Beloborodov2001,Beloborodov2003,Beloborodov2004,Beloborodov2005,Beloborodov2007,Skrzynski_2002,Biagini_2005_WL}, in that we work in granular real space as opposed to lattice momentum space. Only after we have summed over the internal DOFs do we transform to the common picture of lattice momentum space. These ideas closely follow the approach of \cite{Perkins2020}, where standard rules of diagrammatics \cite{AGD} are used with a few additional rules to include granularity into the problem. These new rules are as follows:
\begin{enumerate}
	\item Each electron line receives a grain index, denoted by a Latin character (e.g. the $j$\textsuperscript{th} grain). All internal grain labels are then summed over.
	
	\item Each tunneling vertex, represented by a crossed circle, only allows for nearest neighbor hopping. The tunneling matrix elements are Gaussian distributed \cite{Beloborodov2007}, analogous to the disorder present within a grain, according to
	\begin{equation}
	\begin{split}
		\langle t_{ij}^{\mathbf{k}\mathbf{p}}\rangle&=0, \\
		\langle t_{ij}^{\mathbf{k}\mathbf{p}}t_{lm}^{\mathbf{k}'\mathbf{p}'}\rangle &=
		\begin{cases}
			t^2(\delta_{im}\delta_{jl}+\delta_{il}\delta_{jm})\delta_{\mathbf{k}+\mathbf{k}'=\mathbf{p}+\mathbf{p}'},\\
			0, \qquad \text{otherwise.}
		\end{cases}
	\end{split}
	\end{equation}
	Each pair of correlated tunneling events then carries a factor of $t^{2}a^{d}$, analogous to the homogeneous disorder factor $(2\pi N(0)\tau_{0})^{-1}$, and conserves intragranular momentum.
	
	\item Each current vertex carries a factor of $t_{ij}^{\mathbf{k}\mathbf{p}} e a /\sqrt{\mathcal{N}}$.
\end{enumerate}

The averaged tunneling matrix elements can be related to $\Gamma$ via Fermi's golden rule,
\begin{equation}
	\Gamma = 2\pi N(0)a^{d}t^{2}, \label{Gamma_t_relation}
\end{equation}
where $N(0)$ is the single spin density of states per unit volume at the Fermi surface, and $d$ is the dimensionality of the system.

The electron Green's function for granular systems maintains the same form as in the homogeneous case,
\begin{equation}
	G_{i}(\mathbf{k},i\varepsilon) = \frac{1}{i\varepsilon - \xi_{\mathbf{k}} + \frac{i}{2\tau}\text{sgn}(\varepsilon)}.
	\label{electron_GF}
\end{equation}
However, $\tau^{-1}$ now contains scattering rates due to both impurities within a grain, $\tau_{0}^{-1}$, and tunneling back and forth between nearest neighbor grains,
\begin{equation}
	\frac{1}{\tau} = \frac{1}{\tau_{0}} + z\Gamma, \label{full_scattering_rate}
\end{equation}
where $z$ is the coordination number of the lattice. This is represented by the Dyson series shown in fig. \ref{electron_GF_diagram}, where the thin solid lines denote the free electron Green's functions within a single grain, whilst thick solid lines represent electron Green's functions including impurity scattering and tunneling. The dashed lines between pairs of impurity scattering events and pairs of tunneling events denote their correlation due to averaging over the Gaussian distributions.

\begin{figure}[t]
	\centering
	\includegraphics[width=12.9cm]{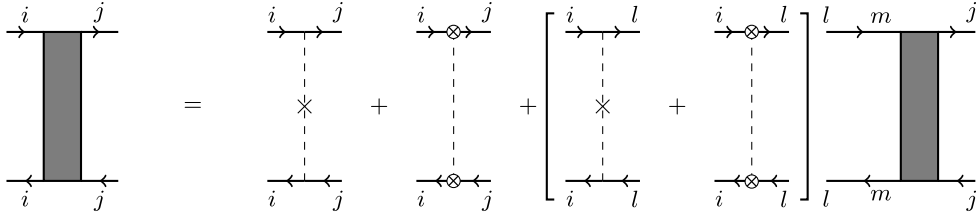}
	\caption{Dyson equation for the granular diffuson.}
	\label{granular_diffuson_diagram}
\end{figure}

We now define the granular diffuson for the system, $\Gamma_{ph}(\mathbf{Q},\mathbf{q},i\varepsilon + i\omega,i\varepsilon)$, via the series shown in fig. \ref{granular_diffuson_diagram}. Under the assumption that an electron scatters several times within a grain before tunneling, $l \ll a$ ($l$ is the elastic mean free path) -- or equivalently $\tau_{0}^{-1} \gg z\Gamma$ and hence $\tau^{-1} \gg z\Gamma$ -- we find the granular diffuson to have the form
\begin{equation}
\begin{split}
	\Gamma_{ph}(\mathbf{Q},\mathbf{q},i\varepsilon + i\omega,i\varepsilon) = \frac{1}{2\pi N(0)\tau^{2}} \frac{\Theta(-\varepsilon(\varepsilon+\omega))}{\mathcal{D}_{0}q^{2} + |\omega| + \Gamma\lambda_{\mathbf{Q}}}.
	\label{granular_diffuson_general}
\end{split}
\end{equation}
In the above, $\lambda_{\mathbf{Q}} = z(1-\gamma_{\mathbf{Q}})$, where $\gamma_{\mathbf{Q}} = z^{-1}\sum_{\alpha}e^{i\mathbf{Q}\cdot\mathbf{a}_{\alpha}}$ is the structure factor, the $\mathbf{a}_{\alpha}$ are the lattice vectors connecting nearest neighbor centres, and the sum is over nearest neighbors. This provides an alternative derivation of the same granular diffuson obtained by Beloborodov et. al. \cite{Beloborodov2007}, in which the internal scattering and external tunneling events are treated on an equal footing. The details are given in the appendix.

The cooperon, $\Gamma_{pp}(\mathbf{Q},\mathbf{q},i\varepsilon + i\omega,i\varepsilon)$, is also obtained in a similar manner, and has the same form as eq. \ref{granular_diffuson_general} with an additional phase breaking rate, $\tau_{\phi}^{-1}$, in the denominator
\begin{equation}
	\Gamma_{pp}(\mathbf{Q},\mathbf{q},i\varepsilon + i\omega,i\varepsilon) = \frac{1}{2\pi N(0)\tau^{2}} \frac{\Theta(-\varepsilon(\varepsilon+\omega))}{\mathcal{D}_{0}q^{2} + |\omega| + \Gamma\lambda_{\mathbf{Q}} + \tau_{\phi}^{-1}}.
	\label{granular_cooperon_general}
\end{equation}
In general we will use capital letters to denote the external momenta, $\mathbf{Q}$, and lowercase letters to denote the internal momenta, $\mathbf{q}$.

From the granular cooperon we now obtain the granular pair propagator,
\begin{equation}
	L(\mathbf{Q},\mathbf{q},i\omega) = -\frac{1}{N(0)}\left[\ln\left(\frac{T}{T_{c,0}}\right) + \psi\left(\frac{1}{2} + \frac{\mathcal{D}_{0}q^{2} + |\omega| + \Gamma\lambda_{\mathbf{Q}} + \tau_{\phi}^{-1}}{4\pi T}\right) - \psi\left(\frac{1}{2}\right)\right]^{-1},
	\label{granular_pair_propagator_general}
\end{equation}
where $\psi(x)$ is the digamma function, $T_{c,0}$ is the bare transition temperature, and $T_{c}$ is the observed transition temperature with suppression due to phase breaking. The latter are related via
\begin{equation}
	\ln\left(\frac{T_{c}}{T_{c,0}}\right) + \psi\left(\frac{1}{2} + \frac{1}{4\pi T_{c}\tau_{\phi,c}}\right) - \psi\left(\frac{1}{2}\right) = 0,
	\label{suppressed_Tc_definition}
\end{equation}
where $\tau_{\phi,c}^{-1}$ is the phase breaking rate at $T_{c}$.

In general, following Lerner at. al. \cite{Lerner}, we will assume that $\Gamma \ll E_{Th} \lesssim T_{c}$, as well as considering the specific sub-case where $E_{Th} \ll T_{c}$. In the small $\Gamma$ limit we may expand the digamma function to obtain,
\begin{equation}
\begin{split}
	L(\mathbf{Q},\mathbf{q},i\omega) = -\frac{1}{N(0)}\Bigg[\ln\left(\frac{T}{T_{c}}\right) &+ \psi\left(\frac{1}{2} + \frac{\mathcal{D}_{0}q^{2} + \tau_{\phi}^{-1}}{4\pi T}\right)  - \psi\left(\frac{1}{2} + \frac{1}{4\pi T_{c}\tau_{\phi,c}}\right) \\
	&\qquad\qquad\quad+ \psi'\left(\frac{1}{2} + \frac{\mathcal{D}_{0}q^{2}+\tau_{\phi}^{-1}}{4\pi T}\right) \frac{|\omega| + \Gamma\lambda_{\mathbf{Q}}}{4\pi T}\Bigg]^{-1}.
	\label{granular_pair_propagator_partial_expansion}
\end{split}
\end{equation}
We write this in a more convenient notation as
\begin{equation}
	L(\mathbf{Q},\mathbf{q},i\omega) = -\frac{1}{N(0)}\left[ \epsilon(\mathbf{q}) + \alpha_{1}(\mathbf{q}) \frac{|\omega| + \Gamma\lambda_{\mathbf{Q}}}{4\pi T}\right]^{-1},
	\label{granular_pair_propagator_partial_expansion_convenient}
\end{equation}
where
\begin{subequations}
	\begin{equation}
		\alpha_{n}(\mathbf{q}) = \psi^{(n)}\left(\frac{1}{2} + \frac{\mathcal{D}_{0}q^{2}+\tau_{\phi}^{-1}}{4\pi T}\right),
		\label{digamma_shorthand}
	\end{equation}
	\text{and}
	\begin{equation}
		\epsilon(\mathbf{q}) = \ln\left(\frac{T}{T_{c}}\right) + \alpha_{0}(\mathbf{q}) - \alpha_{0,c}(\mathbf{0}).
		\label{log_shorthand}
	\end{equation}
	\label{shorthand}%
\end{subequations}
Here we have used the subscript $c$ on $\alpha_{0,c}(\mathbf{0})$ to denote that it is evaluated at the transition temperature. It is worth noting that $\epsilon(\mathbf{0}) \simeq \eta = (T-T_{c})/T_{c}$ when $\eta \ll 1$ and the phase breaking rate is small. This is true in all regions of behavior, except the extreme part of the far-from-$T_{c}$ region where $T$ approaches $2T_{c}$.

Moving onto the sub-case, $E_{Th} \ll T_{c}$, we may further expand the denominator of the pair propagator to yield
\begin{equation}
	L(\mathbf{Q},\mathbf{q},i\omega) = -\frac{1}{N(0)}\Bigg[\epsilon(\mathbf{0}) + \alpha_{1}(\mathbf{q}) \frac{|\omega| + \mathcal{D}_{0}q^{2} + \Gamma\lambda_{\mathbf{Q}}}{4\pi T}\Bigg]^{-1}.
	\label{granular_pair_propagator_full_expansion}
\end{equation}
As in previous literature \cite{Beloborodov2007}, we assume the grains form a cubic lattice of side length $a$, so that
\begin{equation}
	\lambda_{\mathbf{Q}} = 2\sum_{\alpha=1}^{d} \left[1 - \cos(Q_{\alpha}a)\right].
\label{simple_cubic_lambda}
\end{equation}
Considering small $Q$, we see that $\Gamma\lambda_{\mathbf{Q}} \simeq \Gamma a^{2}Q^{2}$, allowing us to identify $\mathcal{D}_{T} = \Gamma a^{2}$ as the granular diffusion coefficient.

\begin{figure}[t]
	\centering
	\includegraphics[width=8.6cm]{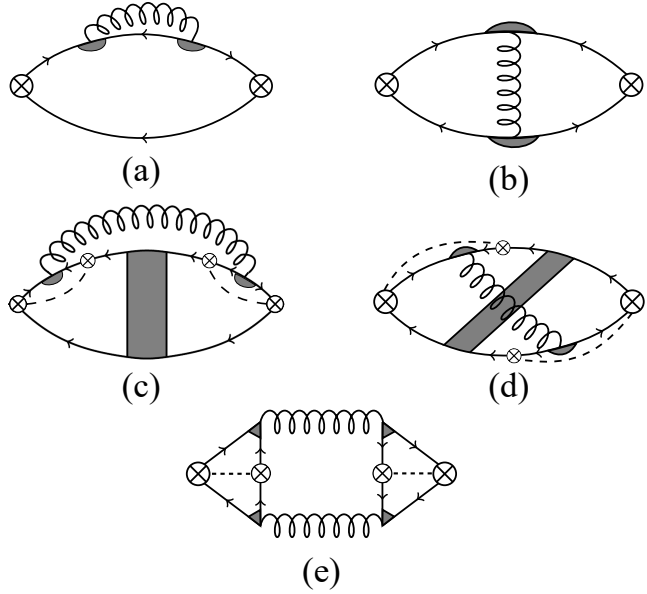}
	\caption{Leading order contributions to the electrical conductivity due to superconducting fluctuations in granular systems. (a) and (c) are density of states (DOS) diagrams; (b) and (d) are Maki-Thompson (MT) diagrams; (e) is the Aslamazov-Larkin (AL) diagram. Diagrams (c) and (d) cancel exactly in the DC conductivity limit, and so are not considered.}
	\label{full_diagram_set}
\end{figure}

Finally, we note that the internal momenta are quantized according to the Neumann boundary condition corresponding to zero current normal to a grain's surface,
\begin{equation}
	\hat{\mathbf{n}} \cdot \nabla \varphi = 0,
	\label{Neumann_BC}
\end{equation}
where $\varphi$ is the electron wavefunction. We shall assume, without loss of generality, that the grains are cubic with side length $a$, so that the internal momenta are quantized as
\begin{equation}
	\mathbf{q} = \frac{\pi}{a}(n_{x},n_{y},n_{z}),
	\label{q_quantisation}
\end{equation}
where $n_{i} = 0,1,2,...$. Clearly the above is for a three dimensional system, but the $d$ dimensional equivalent is trivial to deduce.

Since we have quantized our internal momenta according to eq. \ref{q_quantisation}, we may analyze the singular nature of the pair propagator to understand the existence of three regions of behavior as one approaches $T_{c}$ from above. We begin by identifying the physical energy scales in $L(\mathbf{Q},\mathbf{q},i\omega)$ as $\mathcal{D}_{0}q^{2}$ and $\Gamma\lambda_{\mathbf{Q}}$, which are associated with the internal and external DOFs respectively. These appear as the dimensionless combinations $\mathcal{D}_{0}q^{2}/T$ and $\Gamma\lambda_{\mathbf{Q}}/T$. The smallest non-zero value of $\mathcal{D}_{0}q^{2}$ is $E_{Th}$, so the dimensionless internal energy scale is $E_{Th}/T_{c}$. Similarly, $\lambda_{\mathbf{Q}}$ is, at most, of order unity, so $\Gamma/T_{c}$ is the dimensionless external energy scale.

Looking at the most singular contributions that arise from the pair propagator, we see that when $\epsilon(\mathbf{0}) \simeq \eta \ll \Gamma/T_{c}$, any non-zero $\mathbf{q}$ gives a significantly less singular function. However, we may consider small non-zero $\mathbf{Q}$ which give rise to small changes in the denominator of eq. \ref{granular_pair_propagator_full_expansion}, that are equally singular to the zero lattice momentum piece. We may therefore treat $\mathbf{Q}$ as being continuous in this region. It follows that only the external DOFs are physically relevant here, and so the system appears to be $d$ dimensional. This is the \textit{close-to-$T_{c}$ region}.

Next consider $\Gamma/T_{c} \ll \eta \ll E_{Th}/T_{c}$, where again any non-zero $\mathbf{q}$ leads to less singular contributions, and hence the internal DOFs again play no role here. However, no choice of $\mathbf{Q}$ will generate a notable change in the propagator, so that the external DOFs are effectively unseen in this region. Therefore, neither the internal or external DOFs are physically relevant in this regime, and so the system appears to be quasi-zero dimensional. This is the \textit{intermediate region}.

Finally consider $E_{Th} \ll \eta \leq 1$. As in the intermediate case, the external DOFs are effectively unseen. However, non-zero $\mathbf{q}$ can give equally singular contributions to the zero momentum term. The internal momenta can then be treated as continuous, leading to the system appearing to be $d$ dimensional again. This is the \textit{far-from-$T_{c}$ region}.

The above gives a concrete definition of the dimensional crossovers described in the introduction and the works of Lerner et. al. \cite{Lerner} and Klemencic et. al. \cite{Klemencic2017}. Now that we have established the key ideas behind the granular diagrammatic method, we proceed to calculate the corrections to the electrical conductivity due to superconducting fluctuations.

\section{General fluctuation corrections} \label{Corrections_sec}

By analogy to the homogeneous system, the complete set of diagrams describing the leading order fluctuation corrections to conductivity are shown in fig. \ref{full_diagram_set} \cite{Fluctuations_book}. As we are calculating the electromagnetic response function, $K_{\alpha\beta}(i\Omega) = \Omega\sigma_{\alpha\beta}(i\Omega)$, where $\Omega$ is a bosonic Matsubara frequency, the diagrams should cancel when $\Omega = 0$. We find that this is the case working in the \textit{granular diffusive limit}, $\mathcal{D}_{T}Q^{2} \ll z\Gamma$, which is equivalent to $Q \ll a^{-1}$. In the following we therefore replace $\Gamma\lambda_{\mathbf{Q}}$ with $\mathcal{D}_{T}Q^{2}$.

The new features arising in the granular diagrams of fig. \ref{full_diagram_set} compared to the homogeneous diagrams are the correlated fourth order tunneling events shown in figs. \ref{full_diagram_set}c, \ref{full_diagram_set}d, and \ref{full_diagram_set}e. These are the leading order contributions at $\mathcal{O}(t^{4})$ after averaging over the tunneling matrix elements. Other choices of event placement or correlation pairing either contribute nothing or generate higher order corrections.

We now focus on the explicit calculation of the density of states (DOS), Aslamazov-Larkin (AL), and Maki-Thompson (MT) diagrams, in that order. We assume that the system is in the \textit{granular metallic limit}, $\delta \ll \Gamma \ll E_{Th} \ll T_{c}$, so that $1 \ll g_{T} \ll g_{g}$. We first consider the DOS contribution in detail, constructing the electromagnetic response function in lattice real space, transforming to lattice momentum space, and treating both the internal and external DOFs simultaneously. This yields the most general form of the DOS contribution, after which we begin to consider the three limiting regimes that occur naturally in the pair propagator: $\eta \ll \Gamma/T_{c}$, $\Gamma/T_{c} \ll \eta \ll E_{Th}/T_{c}$, and $E_{Th}/T_{c} \ll \eta \ll 1$. 
We then summarise the expected temperature dependence of the DOS correction to the electrical conductivity deep inside each temperature region.

We then move on to the AL and MT diagrams in turn, presenting fewer mathematical details here, as the same ideas and methods are used as in the DOS calculation. In each case we first present the most general form of the response function with both sets of DOFs, before considering the different limiting behaviors. We finally summarise the temperature dependences of all corrections in each region. A complete set of our results is presented in table \ref{results_table} at the end of this section.

\subsection{DOS corrections}
To the diagram in fig. \ref{full_diagram_set}a we associate the electromagnetic response function
\begin{equation}
	\begin{split}
	K_{\alpha\alpha}(i\Omega) = -\frac{4e^{2}a^{2}t^{2}T^{2}}{a^{2d}\mathcal{N}} &\sum_{\substack{i,l, \\ m}} \sum_{\varepsilon,\omega} \sum_{\substack{\mathbf{k},\mathbf{p}, \\ \mathbf{q}}} \bigg\{ G_{i}(\mathbf{k},i\varepsilon+i\Omega)^{2}G_{i}(\mathbf{q}-\mathbf{k},i\omega-i\varepsilon-i\Omega) \\
	&\times [G_{i+\alpha}(\mathbf{p},i\varepsilon) + G_{i-\alpha}(\mathbf{p},i\varepsilon)] L_{lm}(\mathbf{q},i\omega) \\
	&\times C_{il}(\mathbf{q},i\varepsilon+i\Omega,i\omega-i\varepsilon-i\Omega) C_{im}(\mathbf{q},i\varepsilon+i\Omega,i\omega-i\varepsilon-i\Omega) \bigg\},
	\end{split}
	\label{DOS_LRF}
\end{equation}
where $\varepsilon$ is a fermionic Matsubara frequency, whilst $\Omega$ and $\omega$ are bosonic Matsubara frequencies. The functions $C_{ij}(\mathbf{q},i\varepsilon+i\omega,i\varepsilon) = 2\pi N(0)\tau\,\Gamma_{pp,ij}(\mathbf{q},i\varepsilon+i\omega,i\varepsilon)$ are the closed cooperons at either end of the pair propagator. The labels $i$ and $i\pm\alpha$ belong to the electron propagators, whilst $l$ and $m$ arise from the closed cooperons allowing for tunneling to a new grain, from which the pair propagator begins or ends. The associated electron lines are not seen explicitly in the diagrams, however, as their Green's functions have already been summed over in deriving the closed cooperon.

These cooperons lead to two possible sign choices for the frequencies $\varepsilon$ and $\omega-\varepsilon-\Omega$. We first focus on the case where $\varepsilon+\Omega > 0$, $\omega-\varepsilon-\Omega < 0$, and $\varepsilon < 0$. Before obtaining the conductivity from $K(i\Omega)$, we need to move to the lattice momentum picture. We start by performing the intragranular momentum sums of the electron Green's functions at the Fermi surface, leaving the sum over the small intragranular momenta, $\mathbf{q}$. We then introduce the granular spatial Fourier transforms of the remaining functions via,
\begin{equation}
	L_{ij}(\mathbf{q},i\omega) = \frac{1}{\mathcal{N}}\sum_{\mathbf{Q}}L(\mathbf{Q},\mathbf{q},i\omega)e^{i\mathbf{Q}\cdot\mathbf{R}_{ij}},
	\label{granular_spatial_FT}
\end{equation}
where $\mathbf{R}_{ij} = \mathbf{R}_{i} - \mathbf{R}_{j}$ is the relative position vector connecting the centres of grains $i$ and $j$, located at $\mathbf{R}_{i}$ and $\mathbf{R}_{j}$ respectively.

Now we perform analytic continuation of the Matsubara frequency $i\Omega$ to real frequency $\Omega + i\delta$, where $\delta$ is a positive infinitesimal, to find the retarded electromagnetic response function, $K^{R}_{\alpha\beta}(\Omega)$. The conductivity tensor is then found using
\begin{equation}
	K^{R}_{\alpha\beta}(\Omega) = -i\Omega \sigma_{\alpha\beta}(\Omega).
	\label{Linear_response_conductivity_relation}
\end{equation}
Given that the $\mathcal{O}(\Omega^{0})$ terms cancel, we expand $K^{R}_{\alpha\beta}(\Omega)$ to $\mathcal{O}(\Omega)$ to find the DC conductivity corrections. We further simplify the problem by only retaining the $\omega = 0$ piece of the Matsubara sum to neglect dynamical effects, as in the homogeneous calculation. Doing this leads to
\begin{equation}
	\sigma_{DOS}^{(1)} = \frac{N(0)\Gamma a^{2}e^{2}}{4\pi^{2}Ta^{d}\mathcal{N}} \sum_{\mathbf{Q}}\sum_{\mathbf{q}}
	\psi''\left(\frac{1}{2} + \frac{\mathcal{D}_{0}q^{2}+\Gamma\lambda_{\mathbf{Q}}+\tau_{\phi}^{-1}}{4\pi T}\right) L(\mathbf{Q},\mathbf{q},0).
	\label{sigma_DOS_start}
\end{equation}
Given that $\Gamma \ll T_{c}$, we may neglect the $\Gamma\lambda_{\mathbf{Q}}$ term appearing in the digamma derivative, $\psi''(x)$.

We may replace the $\mathbf{Q}$ sum by an integral, use the granular diffusive limit to replace all occurrences of $\Gamma\lambda_{\mathbf{Q}}$ with $\Gamma a^{2}Q^{2}$ in the integrand, and take the upper limit to be $1/a$. At this point we focus on the $d = 3$ case in order to compare to the experimental results of \cite{Klemencic2017}. We are thus left with
\begin{equation}
	\sigma_{DOS}^{(1)} = \frac{\Gamma e^{2}}{8\pi^{4}Ta}\sum_{\mathbf{q}} \int_{0}^{1} dQ \frac{\alpha_{2}(\mathbf{q}) Q^{2}}{\epsilon(\mathbf{q}) + \frac{\alpha_{1}(\mathbf{q})\Gamma}{4\pi T}Q^{2}}.
	\label{sigma_DOS_Q_integral}
\end{equation}

We note that the alternative sign choice, $\varepsilon+\Omega < 0$, $\omega-\varepsilon-\Omega > 0$, and $\varepsilon < 0$, produces an identical contribution, and so we find the correction due to the DOS to be twice that of eq. \ref{sigma_DOS_Q_integral},
\begin{equation}
	\sigma_{DOS} = \frac{e^{2}}{\pi^{3}a} \sum_{\mathbf{q}}\frac{\alpha_{2}(\mathbf{q})}{\alpha_{1}(\mathbf{q})} \left[1 - \sqrt{\frac{4\pi T\epsilon(\mathbf{q})}{\Gamma\alpha_{1}(\mathbf{q})}}\arctan\left(\sqrt{\frac{\Gamma\alpha_{1}(\mathbf{q})}
	{4\pi T\epsilon(\mathbf{q})}}\right)\right].
	\label{sigma_DOS}
\end{equation}
This is as much progress as can be made with exact analytics. We now consider the different temperature regions by taking the appropriate limits of eq. \ref{sigma_DOS}.

In the close-to-$T_{c}$ region, $\eta \ll \Gamma/T_{c}$, only the $\mathbf{q} = \mathbf{0}$ component gives any significant contribution. All non-zero internal momentum contributions are less singular than the zero momentum piece due to the quantized nature of $\mathbf{q}$ and the fact that the Thouless energy is much larger than the tunneling rate. We therefore reproduce the expected behavior analogous to homogeneous case,
\begin{equation}
	\sigma_{DOS} = \frac{e^{2}}{\pi^{3}a} \frac{\alpha_{2}(\mathbf{0})}{\alpha_{1}(\mathbf{0})} \left[1 - \sqrt{\frac{4\pi T\eta}{\Gamma\alpha_{1}(\mathbf{0})}}\arctan\left(\sqrt{\frac{\Gamma\alpha_{1}(\mathbf{0})}{4\pi T\eta}}\right)\right],
	\label{sigma_DOS_close}
\end{equation}
which is approximately constant for $\eta \ll \Gamma/T_{c}$ \cite{Technical_note}. In this region the characteristic size of a fluctuating Cooper pair is much larger than the typical size of a grain. The granular system thus appears to be a homogeneous disordered medium, with the tunneling events acting as the source of disorder.

To obtain the correction in the intermediate region, $\Gamma/T_{c} \ll \eta \ll E_{Th}/T_{c}$, we again note that only the $\mathbf{q} = \mathbf{0}$ term gives any significant contribution, so that eq. \ref{sigma_DOS_close} is still valid. The argument of the $\arctan$ is now small, so we may expand this to leading order to obtain
\begin{equation}
	\sigma_{DOS} = \frac{\alpha_{2}(\mathbf{0})}{12\pi^{4}} \frac{\Gamma}{T} \frac{e^{2}}{a} \frac{1}{\eta}.
	\label{sigma_DOS_intermediate}
\end{equation}

In the far-from-$T_{c}$ region, $E_{Th}/T_{c} \ll \eta \ll 1$, we need to consider the $\mathbf{q} \neq \mathbf{0}$ terms. Since $\eta \gg E_{Th}/T_{c}$, the summand will be slowly varying for small $\mathbf{q}$ values, and so we may approximate the sum by an integral with an appropriate upper cut-off at $q = q_{c}$. A natural cut-off arises from the diffusive limit, so it appears that $\mathcal{D}_{0}q_{c}^{2} = \tau^{-1}$. However, we need to remember that the digamma derivatives decay quickly and so become small when $\mathcal{D}_{0}q^{2} \geq 4\pi T$, and so we define the cut-off via $\mathcal{D}_{0}q_{c}^{2} = 4\pi T$. As the most singular behavior occurs for $\mathcal{D}_{0}q^{2} \ll 4\pi T$, we expand the digamma functions and their derivatives so that
\begin{subequations}
	\begin{equation}
	\alpha_{n}(\mathbf{q}) \simeq \alpha_{n}(\mathbf{0}) + \frac{\mathcal{D}_{0}q^{2}}{4\pi T}\alpha_{n+1}(\mathbf{0}),
	\label{approx_digamma_shorthand}
	\end{equation}
	\begin{equation}
	\epsilon(\mathbf{q}) \simeq \ln\left(\frac{T}{T_{c}}\right) + \alpha_{0}(\mathbf{0}) - \alpha_{0,c}(\mathbf{0}) + \frac{\alpha_{1}(\mathbf{0})}{4\pi T}\mathcal{D}_{0}q^{2}.
	\label{approx_log_shortand}
	\end{equation}
	\label{approx_shorthand}%
\end{subequations}
Before performing the $\mathbf{q}$ integral, we are able to expand the $\arctan$ again, noting that $\epsilon(\mathbf{q}) \gg \Gamma/T_{c}$ in this case. We therefore arrive at the integral,
\begin{equation}
	\sigma_{DOS} = \frac{\alpha_{2}(\mathbf{0})\Gamma a^{2} e^{2}}{24\pi^{6}T} \int_{0}^{q_{c}} dq \frac{q^{2}}{\eta + \frac{\alpha_{1}(\mathbf{0})}{4\pi T} \mathcal{D}_{0} q^{2}},
	\label{sigma_DOS_far_integral}
\end{equation}
where $q_{c} = \sqrt{4\pi T/E_{Th}}$, and we noted that $\ln(T/T_{c}) \simeq \eta$ when $\eta \ll 1$. This leads to the result
\begin{equation}
	\sigma_{DOS} = \frac{\alpha_{2}(\mathbf{0})}{3\alpha_{1}(\mathbf{0})} \frac{e^{2}}{a} \sqrt{\frac{\Gamma^{2}T}{\pi^{9} E_{Th}^{3}}} \left[1 - \sqrt{\frac{\eta}{\alpha_{1}(\mathbf{0})}}\arctan\left(\sqrt{\frac{\alpha_{1}(\mathbf{0})}{\eta}}\right)\right].
\label{sigma_DOS_far}
\end{equation}
This correction has the same form as the close-to-$T_{c}$ region, albeit with a few changes to the constants appearing alongside $\eta$. We see that for $\eta \ll 1$, the DOS contribution will again be approximately constant. This behavior can be attributed to the fluctuating Cooper pairs being much smaller than the typical grain size, and hence the system again appears homogeneous.

We have therefore shown that there are three temperature regions with different power law relations between $\sigma_{DOS}$ and $\eta$,
\begin{equation}
	\sigma_{DOS} \sim -\frac{e^{2}}{a}
	\begin{cases}
		const.\,, \quad &\eta \ll \frac{\Gamma}{T_{c}} \\[5pt]
		\frac{\Gamma}{T}\eta^{-1}\,, \quad &\frac{\Gamma}{T_{c}} \ll \eta \ll \frac{E_{Th}}{T_{c}} \\[5pt]
		const. \times \sqrt{\frac{\Gamma^{2}T}{E_{Th}^{3}}}\,, \quad &\frac{E_{Th}}{T_{c}} \ll \eta \ll 1.
	\end{cases}
	\label{sigma_DOS_regime_relations}
\end{equation}

\subsection{Aslamazov-Larkin corrections}
The Aslamazov-Larkin diagram of fig. \ref{full_diagram_set}e has the electromagnetic response function,
\begin{equation}
	K_{\alpha\beta}(i\Omega) = -\frac{4T}{a^{d}\mathcal{N}} \sum_{\substack{l,m, \\ 	n,s}}\sum_{\mathbf{q}}\sum_{\omega}\bigg[\widetilde{B}_{\alpha,ls}(\mathbf{q},i\omega) \widetilde{B}_{\beta,nm}(\mathbf{q},i\omega)L_{ml}(\mathbf{q},i\omega+i\Omega)L_{sn}(\mathbf{q},i\omega)\bigg],
\label{AL_LRF_real_space}
\end{equation}
where the functions $\widetilde{B}_{\alpha,ls}(\mathbf{q},i\omega)$ represent the triangular blocks either side of the diagram. These may be written explicitly as

\begin{equation}
\begin{split}
	\widetilde{B}_{\alpha,ls}(\mathbf{q},i\omega,&i\Omega) = \frac{Teat^{2}}{a^{d}} \sum_{i}\sum_{\varepsilon}\sum_{\mathbf{k},\mathbf{p}} \Big[ G(\mathbf{k},i\varepsilon+i\Omega) G(\mathbf{q}-\mathbf{k},i\omega-i\varepsilon) G(\mathbf{q}-\mathbf{p},i\omega-i\varepsilon) \\
	&\times G(\mathbf{p},i\varepsilon) C_{li}(\mathbf{q},i\varepsilon,i\omega-\varepsilon) \Big\{ C_{si+\alpha}(\mathbf{q},i\varepsilon,i\omega-i\varepsilon) - C_{si-\alpha}(\mathbf{q},i\varepsilon,i\omega-i\varepsilon) \Big\} \Big].
\end{split}
\label{AL_block_full_real_space}
\end{equation}
Noting that only the pair propagator and cooperons have explicit dependence on the grain indices, we may move easily to the lattice momentum space picture, so that
\begin{equation}
	K_{\alpha\alpha}(i\Omega) = -\frac{4T}{a^{d}\mathcal{N}} \sum_{\mathbf{Q}}\sum_{\mathbf{q}}\sum_{\omega}\bigg[B_{\alpha}(\mathbf{Q},\mathbf{q},i\omega)^{2} L(\mathbf{Q},\mathbf{q},i\omega+i\Omega)L(\mathbf{Q},\mathbf{q},i\omega)\bigg],
	\label{AL_LRF}
\end{equation}
where
\begin{equation}
\begin{split}
	B_{\alpha}(\mathbf{Q},\mathbf{q},i\omega,i\Omega) = \frac{2Teat^{2}}{a^{d}} &\sin(Q_{\alpha}a) \sum_{\varepsilon} \sum_{\mathbf{k},\mathbf{p}} \bigg[ G(\mathbf{k},i\varepsilon+i\Omega) G(\mathbf{q}-\mathbf{k},i\omega-i\varepsilon) \\
	&\qquad\times G(\mathbf{p},i\varepsilon) G(\mathbf{q}-\mathbf{p},i\omega-i\varepsilon) C(\mathbf{Q},\mathbf{q},i\varepsilon,i\omega-i\varepsilon)^{2} \bigg].
\end{split}
\label{AL_block_full}
\end{equation}

As in the homogeneous case \cite{Fluctuations_book,AL_Sov_Phys,*AL_Phys_Lett}, we may take $\Omega = \omega = 0$ inside the blocks to isolate the most singular behavior of the pair propagators. Computing the momentum and frequency sums within the granular diffusive limit yields,
\begin{equation}
	B_{\alpha}(\mathbf{Q},\mathbf{q}) = \frac{ea^{2}\Gamma N(0)}{2\pi T}\psi'\left(\frac{1}{2} + \frac{\mathcal{D}_{0}q^{2}+\tau_{\phi}^{-1}}{4\pi T}\right)Q_{\alpha}.
	\label{AL_block_general_diffusive}
\end{equation}
We analytically continue the $\omega$ sum in the usual manner \cite{Fluctuations_book}, and take the $\mathcal{O}(\Omega)$ term to arrive at,
\begin{equation}
	\sigma_{AL} = \sum_{\mathbf{Q}}\sum_{\mathbf{q}} \frac{B_{\alpha}(\mathbf{Q},\mathbf{q})^{2}}{\pi TNa^{d}} \int_{-\infty}^{+\infty}dz \frac{\text{Im}[L^{R}(\mathbf{Q},\mathbf{q},z)]^{2}}{\sinh^{2}\left(\frac{z}{2T}\right)},
	\label{sigma_AL_frequency_integral}
\end{equation}
where $L^{R}(\mathbf{Q},\mathbf{q},z)$ is the retarded form of the analytically continued pair propagator. By approximating $\sinh[z/(2T)] \simeq z/(2T)$, we perform the $z$ integration, and replacing the $\mathbf{Q}$ sum by an integral, we find
\begin{equation}
\begin{split}
	\sigma_{AL} = \frac{Te^{2}}{6\pi^{2}\Gamma a} \sum_{\mathbf{q}} \Bigg[3\sqrt{\frac{\alpha_{1}(\mathbf{q})\Gamma}{4\pi T\epsilon(\mathbf{q})}} &\arctan\left(\sqrt{\frac{\alpha_{1}(\mathbf{q})\Gamma}{4\pi T\epsilon(\mathbf{q})}}\right) \\
	&\qquad\qquad- \frac{\alpha_{1}(\mathbf{q})\Gamma [5\alpha_{1}(\mathbf{q})\Gamma + 12\pi T\epsilon(\mathbf{q})]}{\left(\alpha_{1}(\mathbf{q})\Gamma + 4\pi T\epsilon(\mathbf{q})\right)^{2}}\Bigg].
	\label{sigma_AL}
\end{split}
\end{equation}
At this point we consider the behavior of $\sigma_{AL}$ in the different temperature regions.

In the close-to-$T_{c}$ region, $\eta \ll \Gamma/T_{c}$, we take the $\mathbf{q} = \mathbf{0}$ term of eq. \ref{sigma_AL} to obtain
\begin{equation}
	\sigma_{AL} = \frac{1}{8}\sqrt{\frac{\alpha_{1}(\mathbf{0})T}{\pi^{3}\Gamma}}\frac{e^{2}}{a}\frac{1}{\eta^{1/2}},
	\label{sigma_AL_close}
\end{equation}
and hence we recover the standard $d = 3$ result.

In the intermediate region, $\Gamma/T_{c} \ll \eta \ll E_{Th}/T_{c}$, we again consider just the $\mathbf{q} = \mathbf{0}$ component, but expand in terms of $\alpha_{1}(\mathbf{0})\Gamma/(4\pi T\eta)$ to obtain the leading order correction. This results in
\begin{equation}
	\sigma_{AL} = \frac{\pi}{1920}\frac{\Gamma^{2}}{T^{2}}\frac{e^{2}}{a}\frac{1}{\eta^{3}},
	\label{sigma_AL_intermediate}
\end{equation}
and we see that the AL term has the quasi-zero dimensional behavior, $\sigma_{AL} \sim \eta^{-3}$, reproducing the result of \cite{Lerner}. This is due to the fluctuating Cooper pairs being of a comparable size to the typical grain.

Finally, for the far-from-$T_{c}$ region, $E_{Th}/T_{c} \ll \eta \ll 1$, we find
\begin{equation}
	\sigma_{AL} = \frac{1}{960} \sqrt{\frac{\alpha_{1}(\mathbf{0})^{3}\Gamma^{4}}{\pi^{9}TE_{Th}^{3}}} \frac{e^{2}}{a} \frac{1}{\eta^{3/2}}.
	\label{sigma_AL_far}
\end{equation}
This is to be expected given the lack of a $q^{2}$ factor in the integrand due to the current vertices.

In review, we see that the AL term also has three distinct regions, each with different dependences upon the reduced temperature,
\begin{equation}
	\sigma_{AL} \sim \frac{e^{2}}{a} \begin{cases}
		\sqrt{\frac{T}{\Gamma}} \, \eta^{-1/2} \,, \quad &\eta \ll \frac{\Gamma}{T_{c}} \\[5pt]
		\frac{\Gamma^{2}}{T^{2}} \, \eta^{-3} \,, \quad &\frac{\Gamma}{T_{c}} \ll \eta \ll \frac{E_{Th}}{T_{c}} \\[5pt]
		\sqrt{\frac{\Gamma^{4}}{TE_{Th}^{3}}} \, \eta^{-3/2} \,, \quad &\frac{E_{Th}}{T_{c}} \ll \eta \ll 1.
	\end{cases}
	\label{sigma_AL_regime_relations}
\end{equation}

\subsection{Maki-Thompson corrections}
Here we are interested in the diagram shown in fig. \ref{full_diagram_set}b. The linear response function written in lattice momentum space for this diagram is

\begin{equation}
\begin{split}
	K_{\alpha\alpha}(i\Omega) = \frac{4e^{2}a^{2}t^{2}T^{2}}{a^{3d}\mathcal{N}} \sum_{\mathbf{Q}} &\sum_{\substack{\mathbf{k},\mathbf{p}, \\ \mathbf{q}}}\sum_{\varepsilon,\omega} \Bigg[ \cos(Q_{\alpha}a) G(\mathbf{k},i\varepsilon+i\Omega)G(\mathbf{q}-\mathbf{k},i\omega-i\varepsilon-i\Omega) \\
	&\times G(\mathbf{p},i\varepsilon) G(\mathbf{q}-\mathbf{p},i\omega-\varepsilon) L(\mathbf{Q},\mathbf{q},i\omega) \\
	&\times C(\mathbf{Q},\mathbf{q},i\varepsilon+i\Omega,i\omega-i\varepsilon-i\Omega) C(\mathbf{Q},\mathbf{q},i\varepsilon,i\omega-i\varepsilon) \Bigg].
\end{split}
\label{MT_LRF}
\end{equation}
Within the granular diffusive limit $\cos(Q_{\alpha}a)$ can be approximated by unity.

There are two types of contribution to the MT term according to the different sign choices of the Matusbara frequencies of the electron Green's functions. The three possible configurations, starting from the bottom left Green's function and moving clockwise in fig. \ref{full_diagram_set}b, are $++--$, $--++$, and $-+-+$. The $++--$ and $--++$ terms together produce a result, $\sigma_{MT}^{(reg1)}$, that is identical to $\sigma_{DOS}/2$; the $-+-+$ term produces a contribution, $\sigma_{MT}^{(reg2)}$ (identical to $\sigma_{DOS}/2$ in the absence of phase breaking and when $E_{Th} \ll T_{c}$, but differs otherwise), and a more singular (anomalous) piece, $\sigma_{MT}^{(an)}$.

Focusing on the $\omega = 0$ component, the second regular (non-anomalous) part can be written as,
\begin{equation}
	\sigma_{MT}^{(reg2)} = \frac{N(0)\Gamma}{\pi\mathcal{N}} \frac{e^{2}a^{2}}{a^{d}} \sum_{\mathbf{Q}} \sum_{\mathbf{q}} \frac{\alpha_{2}(\mathbf{q})\Gamma\lambda_{\mathbf{Q}}L(\mathbf{Q},\mathbf{q},0)}{\mathcal{D}_{0}q^{2} + \Gamma\lambda_{\mathbf{Q}} + \tau_{\phi}^{-1}},
	\label{sigma_reg2_no_q_expansion}
\end{equation}
where we have not yet assumed anything about the size of $E_{Th}$. Considering $d = 3$, we perform the $\mathbf{Q}$ integral to yield,
\begin{equation}
\begin{split}
	\sigma_{MT}^{(reg2)} &= S_{\phi} + \frac{1}{2\pi^{3}} \frac{e^{2}}{a} \sum_{\mathbf{q}} \frac{\alpha_{2}(\mathbf{q})}{\alpha_{1}(\mathbf{q})} \\
	&\!\!\!\!\times \Bigg[ 1 - \frac{\alpha_{1}(\mathbf{q}) \Gamma\tau_{\phi}}{4\pi T\tau_{\phi}\epsilon(\mathbf{q}) - \alpha_{1}(\mathbf{q})(1+\mathcal{D}_{0}q^{2}\tau_{\phi})} \left(\frac{4\pi T\epsilon(\mathbf{q})}{\alpha_{1}(\mathbf{q})\Gamma}\right)^{3/2} \arctan\left(\sqrt{\frac{\alpha_{1}(\mathbf{q})\Gamma}{4\pi T\epsilon(\mathbf{q})}}\right) \Bigg],
	\label{sigma_reg2_no_q_expansion_Q_integral_result}
\end{split}
\end{equation}
where
\begin{equation}
\begin{split}
	S_{\phi} = \frac{1}{2\pi^{3}} \frac{e^{2}}{a} \sum_{\mathbf{q}} \Bigg[ &\frac{\alpha_{2}(\mathbf{q}) \Gamma\tau_{\phi}}{4\pi T\tau_{\phi}\epsilon(\mathbf{q}) - \alpha_{1}(\mathbf{q})(1+\mathcal{D}_{0}q^{2}\tau_{\phi})} \\
	&\qquad\qquad\qquad\times \left(\frac{1+\mathcal{D}_{0}q^{2}\tau_{\phi}}{\Gamma\tau_{\phi}}\right)^{3/2} \arctan\left(\sqrt{\frac{\Gamma\tau_{\phi}}{1+\mathcal{D}_{0}q^{2}\tau_{\phi}}}\right) \Bigg].
	\label{Phase_coherent_term_shorthand}
\end{split}
\end{equation}
In the absence of phase breaking, $S_{\phi} = 0$ and eq. \ref{sigma_reg2_no_q_expansion_Q_integral_result} collapses to half of eq. \ref{sigma_DOS} when $E_{Th} \ll T_{c}$, as expected. The major new feature we see in the granular $\sigma_{MT}^{(reg2)}$ contribution is the presence of $S_{\phi}$.

The close-to-$T_{c}$ behavior is given by the $\mathbf{q} = \mathbf{0}$ piece of eq. \ref{sigma_reg2_no_q_expansion_Q_integral_result}. To find the intermediate behavior, we expand the $\arctan$ containing $\eta$ in eq. \ref{sigma_reg2_no_q_expansion_Q_integral_result} to third order in its argument, and then take $\mathbf{q} = \mathbf{0}$,
\begin{equation}
	\sigma_{MT}^{(reg2)} = \frac{1}{2\pi^{3}} \frac{\alpha_{2}(\mathbf{0})}{\alpha_{1}(\mathbf{0})} \frac{e^{2}}{a} \left[ 1 + \frac{\alpha_{1}(\mathbf{0})\Gamma\tau_{\phi}}{4\pi T\tau_{\phi}\eta - \alpha_{1}(\mathbf{0})} \left( \frac{1}{3} - \frac{4\pi T\eta}{\alpha_{1}(\mathbf{0})\Gamma} + \frac{\arctan(\sqrt{\Gamma\tau_{\phi}})}{(\Gamma\tau_{\phi})^{3/2}} \right) \right].
	\label{MT_reg2_close}
\end{equation}

Finally, the far-from-$T_{c}$ behavior is found by performing the $\mathbf{q}$ sum numerically in the case that we are unable to perform an expansion in $\mathcal{D}_{0}q^{2}$. If, however, $E_{Th}$ is sufficiently small to allow this expansion, instead of taking the $\mathbf{q} = \mathbf{0}$ piece to obtain the intermediate result, we may expand $\epsilon(\mathbf{q})$ to first order in $\mathcal{D}_{0}q^{2}$, whilst setting $\mathbf{q} = \mathbf{0}$ in $\alpha_{n}(\mathbf{q})$, and replace the $\mathbf{q}$ sum by an integral. This integral has the usual cut-off, and so yields,
\begin{equation}
\begin{split}
	\sigma_{MT}^{(reg2)} = S_{\phi} + \frac{1}{12\pi^{5}} \frac{\alpha_{2}(\mathbf{0})}{\alpha_{1}(\mathbf{0})} \left(\frac{4\pi T}{E_{Th}}\right)^{3/2} &\frac{1}{4\pi T\tau_{\phi}\eta - \alpha_{1}(\mathbf{0})} \\
	&\times \left\{ \Gamma\tau_{\phi}\eta\left[1 - \sqrt{\eta}\arctan\left(\frac{1}{\sqrt{\eta}}\right)\right] - \alpha_{1}(\mathbf{0}) \right\},
\end{split}
\end{equation}
where $S_{\phi}$ will have to be handled numerically.

It is clear that $\sigma_{MT}^{(reg2)}$ will only be singular in the intermediate region, similar to the DOS contribution. In the close-to-$T_{c}$ and far-from-$T_{c}$ regions, the behavior is approximately constant for small phase breaking rates. For larger phase breaking rates, the temperature dependence of $\tau_{\phi}^{-1}$ may become important.

Moving onto the anomalous contribution, we consider only the $\omega = 0$ component to yield
\begin{equation}
	\sigma_{MT}^{(an)} = \frac{N(0)\Gamma}{\pi \mathcal{N}}\frac{e^{2}a^{2}}{a^{d}} \sum_{\mathbf{Q}}\sum_{\mathbf{q}} \frac{\alpha_{1}(\mathbf{q})L(\mathbf{Q},\mathbf{q},0)}{\mathcal{D}_{0}q^{2} + \Gamma\lambda_{\mathbf{Q}} + \tau_{\phi}^{-1}}.
	\label{sigma_MT_an_sums}
\end{equation}
Taking $d = 3$, we perform the $\mathbf{Q}$ integration to give
\begin{equation}
\begin{split}
	\sigma_{MT}^{(an)} &= \frac{2T\tau_{\phi}}{\pi^{2}} \frac{e^{2}}{a} \sum_{\mathbf{q}} \frac{\alpha_{1}(\mathbf{q})}{\alpha_{1}(\mathbf{q})(\mathcal{D}_{0}q^{2}\tau_{\phi} + 1) - 4\pi T\tau_{\phi}\epsilon(\mathbf{q})} \\
	&\!\!\!\!\times \Bigg[ \sqrt{\frac{\mathcal{D}_{0}q^{2} + \tau_{\phi}^{-1}}{\Gamma}} \arctan\left(\sqrt{\frac{\Gamma}{\mathcal{D}_{0}q^{2} + \tau_{\phi}^{-1}}}\right) - \sqrt{\frac{4\pi T\epsilon(\mathbf{q})}{\alpha_{1}(\mathbf{q})\Gamma}} \arctan\left(\sqrt{\frac{\alpha_{1}(\mathbf{q})\Gamma}{4\pi T\epsilon(\mathbf{q})}}\right) \Bigg].
\end{split}
\label{sigma_MT_an}
\end{equation}
For simplicity we initially set $\tau_{\phi}^{-1} = 0$, and return to the discussion of a non-zero value later.

In the close-to-$T_{c}$ region ($\eta \ll \Gamma/T_{c}$), we recover the analogous $d = 3$ homogeneous result
\begin{equation}
	\sigma_{MT}^{(an)} = \sqrt{\frac{T}{2\pi\Gamma}} \frac{e^{2}}{a} \frac{1}{\eta^{1/2}}.
	\label{sigma_MT_an_close}
\end{equation}
In the intermediate region ($\Gamma/T_{c} \ll \eta \ll E_{Th}/T_{c}$) we find,
\begin{equation}
	\sigma_{MT}^{(an)} = \frac{1}{4\pi} \frac{e^{2}}{a} \frac{1}{\eta},
	\label{sigma_MT_an_intermediate}
\end{equation}
which has the same power law form as the DOS term, but with a prefactor of order unity as opposed to $\Gamma/T_{c}$. Finally, in the far-from-$T_{c}$ region ($E_{Th}/T_{c} \ll \eta \ll 1$), we obtain
\begin{equation}
	\sigma_{MT}^{(an)} = \sqrt{\frac{\Gamma^{2}T}{8\pi E_{Th}^{3}}} \frac{e^{2}}{a} \frac{1}{\eta^{1/2}}.
	\label{sigma_MT_an_far}
\end{equation}

In summary, we see that the Maki-Thompson correction has the following dependence upon the reduced temperature,
\begin{equation}
	\sigma_{MT}^{(an)} \sim \frac{e^{2}}{a} \begin{cases}
		\sqrt{\frac{T}{\Gamma}} \, \eta^{-1/2} \,, \quad &\eta \ll \frac{\Gamma}{T_{c}} \\[5pt]
		\eta^{-1} \,, \quad &\frac{\Gamma}{T_{c}} \ll \eta \ll \frac{E_{Th}}{T_{c}} \\[5pt]
		\sqrt{\frac{\Gamma^{2}T}{E_{Th}^{3}}} \, \eta^{-1/2} \,, \quad &\frac{E_{Th}}{T_{c}} \ll \eta \ll 1.
	\end{cases}
	\label{sigma_MT_an_regime_relations}
\end{equation}

The complete set of the DOS, AL, and MT corrections, and their temperature dependence in the three regions of behavior are summarised in table \ref{results_table}.

\begingroup
\renewcommand{\arraystretch}{2}
\begin{table}[t]
	\caption{Summary of the regional dependences of the DOS, AL, and MT corrections upon $\eta$ and the energy scales $\Gamma$, $E_{Th}$, and $T_{c}$, when $\tau_{\phi}^{-1} = 0\,$K.}
	\label{results_table}
	\begin{ruledtabular}
	\begin{tabular}{cccc}
		Diagram & $\eta \ll \frac{\Gamma}{T_{c}}$ & $\frac{\Gamma}{T_{c}} \ll \eta \ll \frac{E_{Th}}{T_{c}}$ & $\frac{E_{Th}}{T_{c}} \ll \eta \ll 1$ \\
		\hline
		$\sigma_{DOS}$ & $1$ & $\frac{\Gamma}{T} \, \eta^{-1}$ & $\sqrt{\frac{\Gamma^{2}T}{E_{Th}^{3}}}$ \\
		$\sigma_{AL}$ & $\sqrt{\frac{\Gamma}{T}} \, \eta^{-1/2}$ & $\frac{\Gamma^{2}}{T^{2}} \, \eta^{-3}$ & $\frac{\Gamma^{2}}{\sqrt{TE_{Th}^{3}}} \, \eta^{-3/2}$ \\
		$\sigma_{MT}^{(an)}$ & $\sqrt{\frac{\Gamma}{T}} \, \eta^{-1/2}$ & $\eta^{-1}$ & $\sqrt{\frac{\Gamma^{2}T}{E_{Th}^{3}}} \, \eta^{-1/2}$ \\
	\end{tabular}
	\end{ruledtabular}
\end{table}
\endgroup

Having demonstrated the existence of three distinct temperature regions in a granular system, we will now discuss the relative magnitude and temperature dependence of the various contributions, to determine which power law behaviors we expect to see experimentally.

\begin{figure*}[t]
	\centering
	\includegraphics[width=17.2cm]{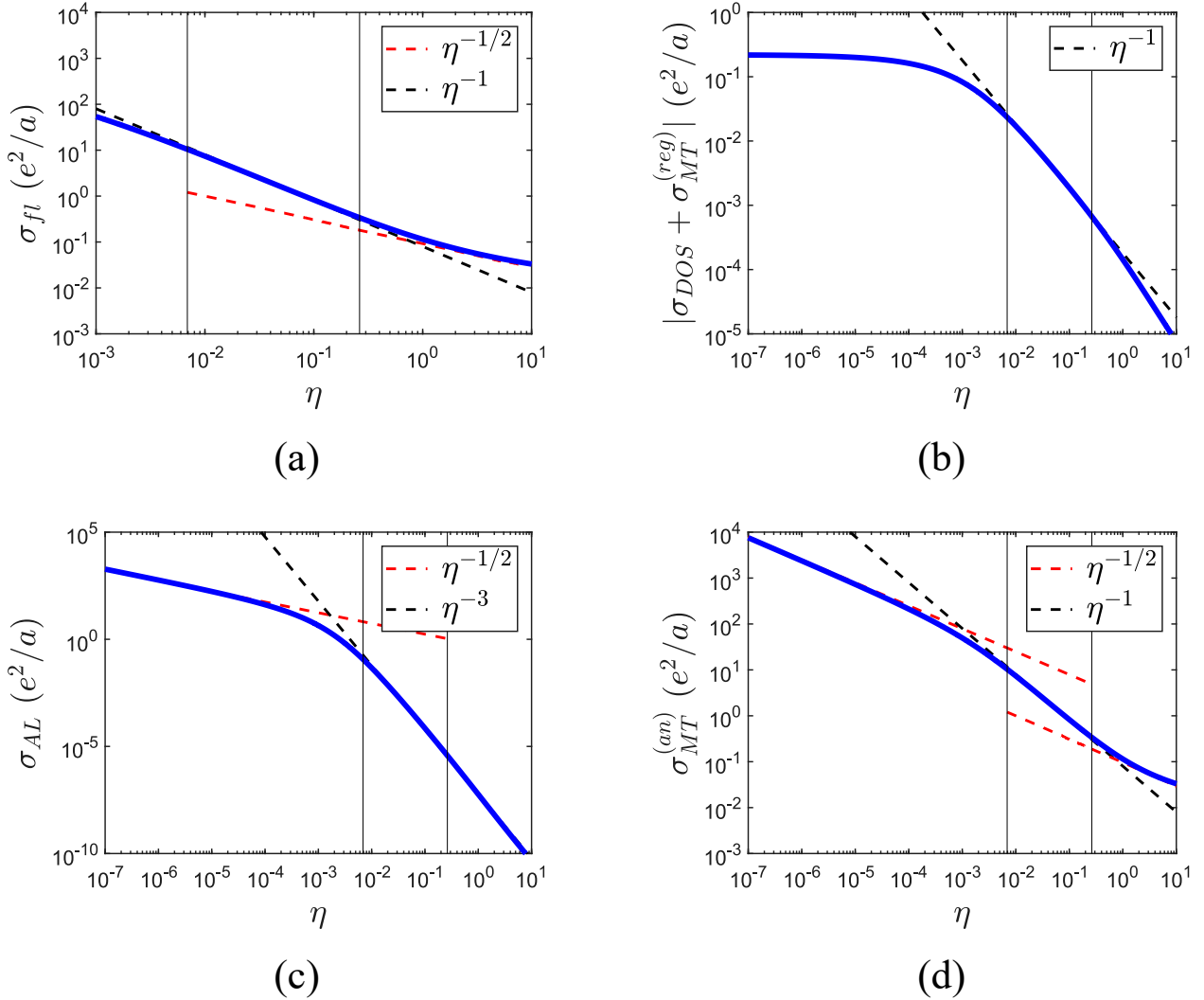}
	\caption{Fluctuation conductivity contributions with no phase breaking. The black vertical lines show where the power law crossovers were seen in Klemencic et. al.'s experiment \cite{Klemencic2017}. The dashed lines act as guides for power law behavior. (a) Total fluctuation conductivity; (b) DOS plus regular MT contribution; (c) AL contribution; (d) anomalous MT contribution.}
	\label{Conductivity_corrections_no_tau_phi}
\end{figure*}

\section{Theoretical discussion} \label{Theoretical_discussion_sec}

Unlike the AL term, the MT and DOS terms have the same power law behavior in both their close-to-$T_{c}$ and far-from-$T_{c}$ regions. This is due to the MT and DOS terms being $\mathcal{O}(t^{2})$, whilst AL is $\mathcal{O}(t^{4})$. The $\mathcal{O}(t^{4})$ behavior results in the generation of a $\sin(Q_{\alpha}a)$ term, and hence an additional factor of $Q$, within each block of the AL diagram, as in the homogeneous case. However, no additional factors of $q$ are generated, so the internal and external DOFs act differently. In contrast, the MT and DOS terms do not gain additional factors of either internal or external momenta. It follows that the $Q$ and $q$ integrals are equivalent in close-to-$T_{c}$ and far-from-$T_{c}$ regions, leading to the same power law behavior.

Examining the magnitude of the diagrams, we see that the AL correction cannot dominate the anomalous MT correction in the absence of a significant phase breaking rate. In the close-to-$T_{c}$ region, $\sigma_{AL} \sim \sigma_{MT}^{(an)}$, whereas $\sigma_{AL} \sim (\Gamma/T)^{2}\eta^{-3}$ and $\sigma_{MT}^{(an)} \sim \eta^{-1}$ in the intermediate region. Despite the AL term being more singular in $\eta$, its prefactor is much smaller since $\Gamma \ll T_{c} \lesssim T$. Simple comparison of these terms shows that the $\eta^{-3}$ behavior of the AL correction can only dominate over the anomalous MT term when $\eta \lesssim \Gamma/T_{c}$, which is clearly not in the region where we expect these power laws to exist. Thus, if $\tau_{\phi}^{-1} = 0$, then all behavior seen in the intermediate regime would be due to the anomalous MT contribution.

The DOS term is of order $\Gamma/T_{c}$, so the AL contribution may be able to dominate over the DOS correction within part of the intermediate regime. The region of AL dominance would depend on the relative size of $\sqrt{\Gamma/T_{c}}$ and $E_{Th}/T_{c}$. In the far-from-$T_{c}$ region, the AL term will only dominate over the DOS when $\eta \lesssim (\Gamma/T_{c})^{2/3}$, which may not occur within this region, again depending upon the size of $(\Gamma/T_{c})^{2/3}$ compared to $E_{Th}/T_{c}$.

Looking at the total fluctuation conductivity, in the absence of phase breaking mechanisms, we therefore expect the anomalous MT and AL terms to dominate in the close-to-$T_{c}$ region, whilst the anomalous MT and DOS contributions will dominate in the intermediate and far-from-$T_{c}$ regions. The resulting fluctuation conductivity would hence be
\begin{equation}
	\sigma_{fl} \sim \frac{e^{2}}{a} \begin{cases}
	\sqrt{\frac{T}{\Gamma}}\frac{1}{\eta^{1/2}} \,, \quad &\eta \ll \frac{\Gamma}{T_{c}} \\[5pt]
	\frac{1}{\eta} \,, \quad &\frac{\Gamma}{T_{c}} \ll \eta \ll \frac{E_{Th}}{T_{c}} \\[5pt]
	\sqrt{\frac{\Gamma^{2}T}{E_{Th}^{3}}}\frac{1}{\eta^{1/2}} \,, \quad &\frac{E_{Th}}{T_{c}} \ll \eta \ll 1.
	\end{cases}
	\label{Expected_fluctuation_conductivity_no_tau_phi}
\end{equation}

Including a phase breaking mechanism, $\tau_{\phi}^{-1} \neq 0$, suppresses the anomalous MT term. If the suppression is large enough, then the anomalous MT and DOS terms may cancel almost perfectly, as they have the same intermediate behavior and are opposite in sign. In this case, the $\eta^{-3}$ power law of the AL term may be able to dominate in the intermediate region and therefore be observable. The exact temperature dependence of the phase breaking mechanisms present in granular systems is currently not known, and is beyond the scope of this paper. We therefore cannot rule out the possibility of a novel phase breaking mechanism that alters the temperature dependence of the anomalous MT term, such that it now produces an $\eta^{-3}$ power law in the intermediate region.

\begin{figure}[t]
	\centering
	\includegraphics[width=8.6cm]{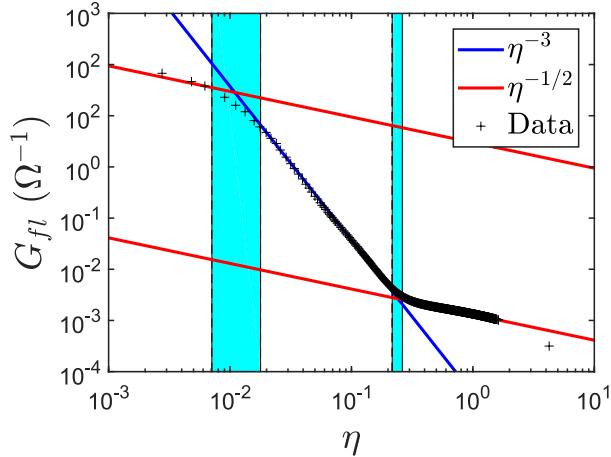}
	\caption{Experimental measurements of the fluctuation conductance from the work of Klemencic et. al. \cite{Klemencic2017} with power law fittings in each region. These measurements were taken on a 329 nm thick BNCD film. The shaded areas show the approximate regions for the two crossovers.}
	\label{Fluctuation_conductivity_experimental_data}
\end{figure}

\section{Comparison to experiment} \label{Comparison_sec}
In what follows, we make use of eqs. \ref{sigma_DOS}, \ref{sigma_AL}, and \ref{sigma_MT_an} with the parameters $T_{c}$, $E_{Th}$, $\Gamma$, $a$, and $\delta$ being taken from the works of Klemencic et. al. \cite{Klemencic2017,Klemencic2019}, and proceed to compute these numerically to make comparison to experimental observation. We take $T_{c} = 3.8\,$K, $E_{Th} = 1\,$K ($\mathcal{D}_{0} = 13.1\text{ cm s}^{-1}$), $\Gamma = 2.62\times10^{-2}\,$K, $a = 10^{-7}\,$m, and $\delta = 5.6 \times 10^{-3}\,$K (this corresponds to a carrier concentration of $n = 10^{27}\text{ m}^{-3}$). With these values we cannot approximate the $\mathbf{q}$ sum as an integral at any point, and thus the sum must be performed explicitly. Due to our assumption of the diffusive limit, $\mathcal{D}_{0}q^{2} \ll \tau_{0}^{-1}$, we need only include the $\mathbf{q} = \mathbf{0}$ term. We therefore do not include the internal DOFs in our following analysis.

In the absence of phase breaking mechanisms, we plot the theoretically expected fluctuation conductivity in fig. \ref{Conductivity_corrections_no_tau_phi}a. Comparing this to the experimental data of Klemencic et. al. \cite{Klemencic2017} in fig. \ref{Fluctuation_conductivity_experimental_data} we see that the anomalous MT and AL contributions dominate in the close-to-$T_{c}$ region, producing the expected $\eta^{-1/2}$ behavior. In the intermediate region, the anomalous MT term overtakes the AL correction leading to an $\eta^{-1}$ power law. Finally, in the far-from-$T_{c}$ region, the anomalous MT term generates an $\eta^{-1/2}$ dependence, whilst dominating the AL contribution. Note that this far-from-$T_{c}$ behavior does not originate from the internal DOFs.

\begin{figure*}[t]
	\centering
	\includegraphics[width=17.2cm]{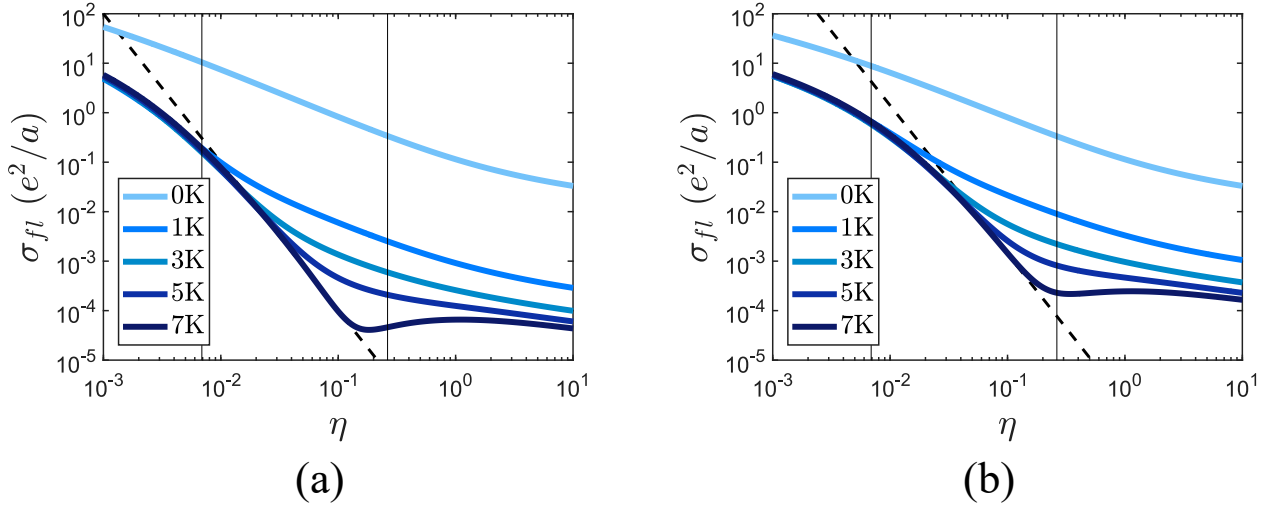}
	\caption{Fluctuation corrections with various constant phase breaking rates, $\tau_{\phi}^{-1}$. The dashed black line acts as a guide for $\eta^{-3}$ behavior. (a) shows the total fluctuation conductivity using parameters based upon Klemencic et. al.'s experiment \cite{Klemencic2017}. (b) shows the total fluctuation conductivity using a custom set of parameters.}
	\label{Conductivity_corrections_varying_tau_phi}
\end{figure*}

This prediction matches the experimental data shown in fig. \ref{Fluctuation_conductivity_experimental_data} in the close-to-$T_{c}$ and far-from-$T_{c}$ regions. However, the data shows a clear $\eta^{-3}$ power law in the intermediate region, which can be attributed to the AL term. This implies that the anomalous MT contribution must experience some form of suppression to allow for the AL power law to dominate.

Let us consider a simple mechanism of suppression in the form of a constant phase breaking rate. In real systems, $\tau_{\phi}^{-1}$ will also have a temperature dependence, but that is beyond the scope of this paper. Phase breaking is often neglected in 3D systems, and only used as a momentum cut-off in the 1D and 2D cases to avoid infrared divergence issues. In reality, however, phase breaking processes exist in all dimensionalities and so we should not neglect them. In fig. \ref{Conductivity_corrections_varying_tau_phi}a we consider the effect of a range of constant phase breaking rates upon the total fluctuation conductivity. For large phase breaking rates, this simple mechanism suppresses the anomalous MT term sufficiently that it almost cancels the DOS correction in the close-to-$T_{c}$ and intermediate regions. Moving further away from the transition, the anomalous MT term begins to deviate from the DOS behavior, and starts to dominate in the far-from-$T_{c}$ region. This constant $\tau_{\phi}^{-1}$ is a simple and not entirely physical mechanism, and leads to a minimum in the fluctuation conductivity when $\tau_{\phi}^{-1} \sim T$.

For comparison, let us consider a different set of physical parameters to try and shift the theoretical close-to-$T_{c}$ to intermediate crossover to higher values of $\eta$. In fig. \ref{Conductivity_corrections_varying_tau_phi}b we choose $\Gamma = 0.1\,$K, and $E_{Th} = 3\,$K. Here we can see that the close-to-$T_{c}$ to intermediate crossover in $\sigma_{fl}$ occurs at a value of $\eta$ nearer to that seen in experiment. The $\eta^{-3}$ power law still requires a large phase breaking rate to appear, and persists over a smaller range of $\eta$. We again observe a minimum occurring in the conductivity for large $\tau_{\phi}^{-1}$.

The minimum in $\sigma_{fl}$ that appears for large $\tau_{\phi}^{-1}$ is extremely small, as is the magnitude of the fluctuation data in the far-from-$T_{c}$ region. As a result, the shape of this data is very sensitive to the fitting applied in the high temperature region ($T \gg T_{c}$). A small change in the fitting parameters for the high temperature data can lead to a significant change in the far-from-$T_{c}$ fluctuation data. Therefore, obtaining the exact $\eta$ dependence of the far-from-$T_{c}$ region is not trivial.

In order to truly compare the theoretical and experimental results, one would need a model of the temperature dependence of $\tau_{\phi}$, which has not been fully addressed in the literature. In particular, one would need to determine how the granularity of a material influences $\tau_{\phi}$ in the fluctuation region. We are left with two possibilities: either the anomalous MT term is suppressed such that the AL term dominates in the intermediate region, or the phase breaking rate has a novel temperature dependence that gives the anomalous MT correction an $\eta^{-3}$ power law. In either case, we can attribute the close-to-$T_{c}$ behavior to the AL and anomalous MT terms, and the far-from-$T_{c}$ behavior to the anomalous MT term. To fully understand the intermediate region, further study of the phase breaking rate in granular systems is required both theoretically and experimentally.

\section{Conclusions} \label{Conclusions_sec}
In this paper we have demonstrated that in a metallic superconducting granular system close to and above $T_{c}$, three regions of behavior exist within the fluctuation conductivity. This was shown by including both internal and external DOFs into the theoretical analysis, to reflect the character of a Cooper pair's varying size compared to the typical grain size. This can be understood in terms of the existence of two coherence lengths in the granular system,
\begin{equation}
	\xi_{g} = \sqrt{\frac{\pi\mathcal{D}_{0}}{8T_{c}\eta}}, \qquad \xi_{T} = \sqrt{\frac{\pi\mathcal{D}_{T}}{8T_{c}\eta}},
\end{equation}
where $\xi_{g}$ is the intragrain coherence length, and $\xi_{T}$ is the intergrain coherence length. These can be obtained from the prefactors of $q^{2}$ and $Q^{2}$, respectively, in the pair propagator of eq. \ref{granular_pair_propagator_full_expansion}. The close-to-$T_{c}$ region occurs when $\xi_{T} \gtrsim a$, the intermediate regime occurs when $\xi_{T} \lesssim a \lesssim \xi_{g}$, and the far-from-$T_{c}$ region occurs when $\xi_{g} \lesssim a$.

In order for the $\eta^{-3}$ power law, which is seen experimentally, to be observable, we found that the inclusion of a significant phase breaking rate was required, $\tau_{\phi}^{-1} \sim T_{c}$ -- such values have been seen experimentally \cite{Meiners-Hagen2001}. This is necessary to suppress the anomalous MT term, such that it cancels almost perfectly with the DOS contribution, thus allowing the AL behavior to dominate. As the phase breaking rate is increased, the observable $\eta^{-3}$ region becomes larger, but eventually an uncharacteristic minimum develops in the fluctuation conductivity. Looking at larger values of $\tau_{\phi}^{-1}$ leads to the DOS term dominating, and hence a negative $\sigma_{fl}$. We note that the far-from-$T_{c}$ fluctuation conductivity extracted in experiment is sensitive to the high temperature fitting, and so a minimum may actually be present in the data. The assumption of a constant phase breaking rate is not realistic, and a detailed understanding of the temperature dependence of $\tau_{\phi}^{-1}$ may be necessary for better fitting of theory to experiment.

\begin{acknowledgments}
	The author D. T. S. Perkins would like to thank Manjinder Kainth and Rose Davies for useful discussions and feedback regarding the contents of this paper and its presentation. D. T. S. Perkins acknowledges funding from the UK Engineering and Physical Sciences Rsearch Council (EPSRC). G. M. Klemencic wishes to acknowledge financial support by the European Research Council under the EU Consolidator Grant ‘SUPERNEMS’ (647471) and the UK EPSRC under the grant `A Diamond Bridge to Phase Slip Physics' (EP/V048457/1).
\end{acknowledgments}

\appendix*
\section{Derivation of the Granular Diffuson}
Here we provide an alternative derivation of the granular diffuson to that of Beloborodov et. al. \cite{Beloborodov2007}, with the inclusion of external DOFs. From the diagram in fig. \ref{granular_diffuson_diagram} the first two terms form the first piece of the series, and may be written respectively as
\begin{equation}
\begin{split}
	S^{(1)}_{ji} &= \frac{1}{2\pi N(0)\tau_{0}}\delta_{ij}, \\
	S^{(2)}_{ji} &= a^{d}t^{2}\delta_{\langle ji \rangle},
	\label{diffuson_first_terms}
\end{split}
\end{equation}
where
\begin{equation}
	\delta_{\langle ji \rangle} = \begin{cases}
		1, \qquad i, \, j \text{ nearest neighbors} \\
		0, \qquad \text{otherwise.}
	\end{cases}
	\label{nearest_neighbor_delta_function}
\end{equation}
We first compute the diffuson self-energy
\begin{equation}
	\Pi_{ml} = \frac{\delta_{ml}}{a^{d}}\sum_{\mathbf{k}} G(\mathbf{k}+\mathbf{q},i\varepsilon+i\omega)G(\mathbf{k},i\varepsilon),
	\label{diffuson_self_energy_sum}
\end{equation}
which is calculated exactly as in the homogeneous case to yield
\begin{equation}
	\Pi_{ml}(\mathbf{q},i\omega) = 2\pi N(0)\tau(1 - \omega\tau - \mathcal{D}_{0}q^{2}\tau)\delta_{ml}.
	\label{diffuson_self_energy}
\end{equation}
The Dyson equation for the diffuson can then be written as,
\begin{equation}
	\Gamma_{ph,ji}(\mathbf{q},i\omega) = S^{(1)}_{ji} + S^{(2)}_{ji} + \sum_{l,m} \Big[S^{(1)}_{li} + S^{(2)}_{li}\Big]\Pi_{ml}(\mathbf{q},i\omega)\Gamma_{ph,jm}(\mathbf{q},i\omega).
	\label{diffuson_dyson_equation}
\end{equation}
To solve this we move to lattice momentum space via the transform in eq. \ref{granular_spatial_FT}, to obtain
\begin{equation}
	\Gamma_{ph}(\mathbf{Q},\mathbf{q},i\omega) = \Big[\Big\{S^{(1)}(\mathbf{Q}) + S^{(2)}(\mathbf{Q})\Big\}^{-1} - \Pi(\mathbf{Q},\mathbf{q},i\omega)\Big]^{-1}.
	\label{diffuson_dyson_equation_lattice_momentum}
\end{equation}
We next rewrite $S^{(1)}_{ji}$ and $S^{(2)}_{ji}$ as
\begin{equation}
	S^{(1)}(\mathbf{Q}) + S^{(2)}(\mathbf{Q}) = \frac{1}{2\pi N(0)\tau_{0}} + a^{d}t^{2}\sum_{\alpha}e^{iQ_{\alpha}a} = \frac{1}{2\pi N(0)\tau}\left[1 + z\Gamma\tau(\gamma_{\mathbf{Q}} - 1)\right],
	\label{diffuson_first_terms_lattice_momentum}
\end{equation}
which may be substituted into eq. \ref{diffuson_dyson_equation_lattice_momentum}.

Finally, we assume that an electron scatters several times within a grain before tunneling to a neighboring grain, so that $z\Gamma\tau_{0} \ll 1$ and hence $z\Gamma\tau \ll 1$. We may now expand the reciprocal of eq. \ref{diffuson_first_terms_lattice_momentum} to leading order in $z\Gamma\tau(\gamma_{\mathbf{Q}} - 1)$, and substitute the result into eq. \ref{diffuson_dyson_equation_lattice_momentum}. This yields the diffuson given in eq. \ref{granular_diffuson_general}.

\bibliography{Fluctuation_Spectroscopy_in_Granular_Superconductors_with_Application_to_Boron-doped_Nanocrystalline_Diamond}

\begin{thebibliography}{22}
\bibitem [{Tec()}]{Technical_note}%
  \BibitemOpen
  \href@noop {} {}\bibinfo {note} {In eq. \ref{sigma_DOS_close}, $\eta$ should
  really be $\epsilon(\mathbf{0})$ -- the replacement of $\epsilon(\mathbf{0})$
  by $\eta$ is only valid when the phase breaking rate is small. We have
  written $\eta$ in this equation as we are initially interested in deriving
  the power law behaviors in the reduced temperature, in the absence of phase
  breaking. The same is true in eqs. \ref{sigma_DOS_intermediate},
  \ref{sigma_AL_close}, \ref{sigma_AL_intermediate}, and
  \ref{MT_reg2_close}.}\BibitemShut {Stop}%
\end{thebibliography}

\end{document}